\documentclass[%
 reprint,
groupedaddress,
 amsmath,amssymb,
 aps,
floatfix,
]{revtex4-1}

\usepackage{graphicx}
\usepackage{dcolumn}
\usepackage{bm}
\usepackage{amsmath}
\usepackage{amssymb}
\usepackage{subfig}
\usepackage{float}

\usepackage{verbatim}
\usepackage{booktabs}
\usepackage{color}

\newcommand{\modif}[1]{{\color{black}#1}}
\begin{document}

\title{Hydrodynamic shocks in  microroller suspensions}%

\author{Blaise Delmotte}
\email{delmotte@courant.nyu.edu}
 \affiliation{Courant Institute of Mathematical Sciences,
New York University, New York, NY 10012, USA.}
\author{Michelle Driscoll}%
\email{mdriscoll@nyu.edu}
\affiliation{%
 Department of Physics, New York University, New York, NY 10003, USA.
}%

\author{Paul Chaikin}
\email{chaikin@nyu.edu}
\affiliation{
 Department of Physics, New York University, New York, NY 10003, USA.
}%

\author{Aleksandar Donev}
\email{donev@courant.nyu.edu}
\affiliation{%
 Courant Institute of Mathematical Sciences,
New York University, New York, NY 10012, USA. 
}%

\date{\today}
\begin{abstract}
We combine experiments, large scale simulations and continuum models to study the emergence of coherent structures  in a suspension of magnetically driven microrollers sedimented near a floor. Collective hydrodynamic effects are predominant in this system, leading to strong density-velocity coupling. We characterize a uniform suspension and show that density waves propagate freely in all directions in a dispersive fashion. When sharp density gradients are introduced in the suspension, we observe the formation of a shock. Unlike Burgers' shock-like structures observed in other active and driven confined hydrodynamic systems, the shock front in our system has a well-defined finite width and moves rapidly compared to the mean suspension velocity. We introduce a continuum model demonstrating that the finite width of the front is due to far-field nonlocal hydrodynamic interactions and governed by a geometric parameter: the average particle height above the floor.  
\end{abstract}

\pacs{Valid PACS appear here}
                    
\maketitle

\begin{figure}
\includegraphics[width=0.9\columnwidth]{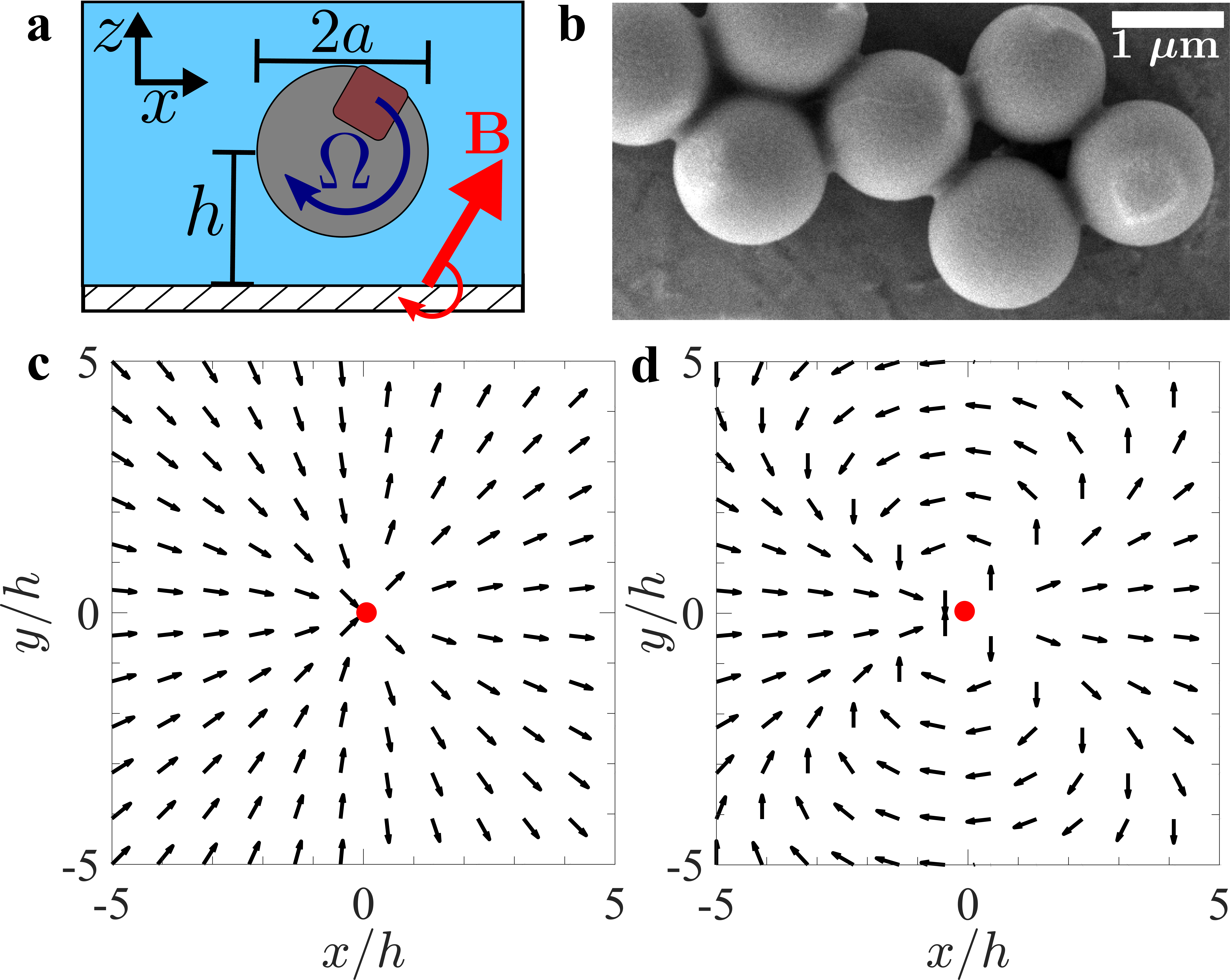}
\caption{\textbf{a} Schematic of a single microroller rotated by a magnetic field $\mathbf{B}$. \textbf{b} SEM image of the microrollers.  \textbf{c} Flow field in the plane parallel to the floor induced by a rotlet rotating about the $y$-axis. \textbf{d} Flow field induced by a potential dipole directed along the $x$-axis. The red circles represents the particle center.}
\label{fig:rotlet_sketch}
\end{figure}

Large-scale  structures can emerge naturally from the dynamics of driven and active systems \cite{Vicsek2012}.  These structures result from the collective, coherent motion of many individual units, and although similar phenomena are seen in widely disparate systems \cite{Marchetti2013, Ramaswamy2010, Saintillan2013}, the interactions that result in collective and coherent motion strongly depend on the specifics of the system being considered. Colloidal suspensions, for example, are always in the Stokes (overdamped) limit due to their small scale.  In this limit, the interactions between the colloidal particles are long-ranged and strongly depend on the presence of nearby boundaries. \modif{Despite the linearity of the equations for the fluid flow in the Stokes regime, elucidating the precise role of hydrodynamic interactions in confined or bounded systems is still an open and challenging problem.}

\modif{Under strong in-plane confinement, i.e.\ in a Hele-Shaw cell, active suspensions exhibit coherent motion at large scales and phase transitions to polar and ordered states.}  For example, recent experiments \cite{Thutupalli2011} and models  \cite{Brotto2013,Lefauve2014,Tsang2014,Tsang2015, Yeo2015, Zottl2014} have shown that hydrodynamic and steric interactions lead to the emergence of collective motion and structure formation in the form of swirls and vortices \cite{Lefauve2014, Tsang2015}, asters \cite{Lefauve2014,Tsang2015}, or polarized density waves \cite{Lefauve2014, Tsang2016, Brotto2013, Caussin2014}.  
In addition to using motile particles, a background flow can also be used to drive a suspension, leading to a rich and diverse array of structure formation: long-ranged  orientational correlations \cite{Shani2014}, density fluctuations at all scales \cite{Desreumaux2013}, and the formation of Burgers-like shocks \cite{Beatus2009, Champagne2011, Lefauve2014,Tsang2016}. \modif{In all of these strongly-confined driven suspensions, despite the difference in propulsion mechanism/driving, the local flow field around a particle is always quasi-two-dimensional (q2D) and can be modeled as a potential dipole  \cite{Cui2004, Beatus2012, Brotto2013}.} Here we show that related but quite different structure formation can emerge from a fundamentally different system, with a different particle-induced flow field and \modif{a different type of confinement}. 
 This contrasts with the prediction \cite{Desreumaux2013} that only dipolar hydrodynamic perturbations can generate such dynamics.

In this work, we investigate the dynamics of microrollers, colloidal particles rotating near a floor. We have previously shown that coherent structures emerge naturally in this system: the microrollers organize into a shock front which then becomes unstable and emits stable motile structures  of a well defined size, called ``critters" \cite{critters}. In this paper, we study in detail the formation of the initial shock front, and density fluctuations to explore how they could lead to the cascade of instabilities observed in this system.
In contrast to previously studied Burgers-like shocks, the shock front we observe in this system has a finite width and propagates much more quickly than the suspension. Using a combined approach of continuum modeling, experiments, and large-scale, 3D numerical simulations, we demonstrate that the origin of this new kind of shock is rooted in the nonlocal hydrodynamic interactions that result from rotational driving.

\begin{figure*}
\includegraphics[width=1.90\columnwidth]{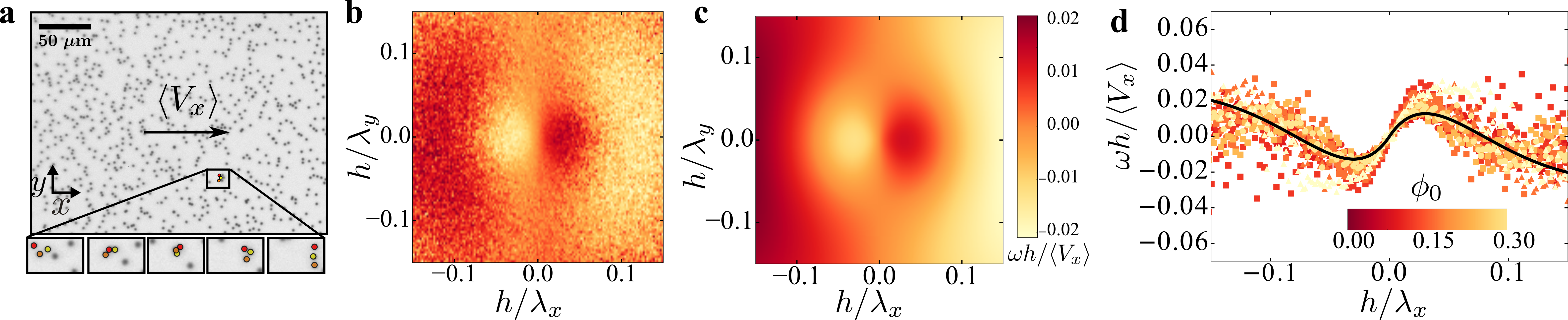}
\caption{\textbf{a} Microroller suspension ($\phi_0 = 0.09$). The magnified inset shows the formation and breakup of an individual cluster (images 1 s apart). \textbf{b} Normalized experimental dispersion curve $\omega h/\langle V_x \rangle$ obtained for $\phi_0 = 0.25$ and $\Omega = 62.8$ rad/s, where the wavenumbers are normalized by $h/2\pi$.  \textbf{c} Normalized theoretical dispersion curve obtained from Eq. \eqref{eq:dispersion_2D}. \textbf{d} Normalized dispersion for $k_y = 0$. Symbols indicate experimental measurements, curve is the theoretical result. Symbol color indicates density, and symbol shape indicates frequency, $\blacksquare: \Omega = 31.4$ rad/s, \textbullet$: \Omega = 62.8$ rad/s, $\blacktriangle: \Omega = 125.7$ rad/s).}
\label{fig:system_kymo}
\end{figure*}

Our system consists of magnetically driven colloidal microrollers, see Fig.\ \ref{fig:rotlet_sketch}a,b. These microrollers are suspended in a sealed chamber of depth $H=$ 200 $\mu$m, width $W=$ 2 mm, length $L=$ 50 mm, and are much denser than the surrounding quiescent fluid. As a result they readily sediment, and remain near the chamber floor, see inset in Fig. \ref{fig:rotlet_sketch}a.  They do not rest on the floor, but are suspended by thermal motion at their gravitational height, $h$, which is set by the balance of thermal energy and their buoyant mass, $h=a+k_BT/mg$, where $a = 0.656$ $\mu$m is the colloid radius, $k_B$ the Boltzmann constant, $T  \approx 298$ K the fluid temperature, $m = 1.27\cdot10^{-15}$ kg their buoyant mass, and $g$ the gravitational acceleration. The gravitational height, $h$, of our microrollers is 1 $\mu$m, which we verified by measuring their translational diffusion constant and comparing with the calculated value. 
In our system, $h$ is two orders of magnitude smaller than the chamber height, $H$, and we only consider particles in a small region (430$\times$430 $\mu$m) in the middle of the chamber, quite far from any lateral wall.  Thus, this closed system is well-approximated by a system with only one boundary (the floor), i.e.\ an infinite half-space.

The magnetic microrollers in the experiments are polymer colloids (3-methacryloxypropyl trimethoxysilane) embedded with a hematite cube and suspended in deionized water \ \cite{Sacanna2012}.  Hematite is a canted antiferromagnet, giving the particles a small permanent moment, $\lvert \mathbf{m}\rvert \sim 5\cdot10^{-16}$ A$\cdot$m$^2$, so that particle motion can be driven by a rotating magnetic field, $\mathbf{B} = B_0\left[\cos(\Omega t)\bm{\hat{x}}+\sin(\Omega t)\bm{\hat{z}}\right]$.  The field is generated using tri-axial coils.  Below a critical frequency, $\Omega_c = 170 $ rad/s, all of the particles rotate synchronously with the applied magnetic field at a rate $\Omega$ \cite{critters}. All experiments were done at frequencies below $\Omega_c$ and at fixed field magnitude, $B_0$ = 2.9 mT.  We emphasize that although the colloidal particles are magnetic, the dominant interparticle interactions in this system are hydrodynamic due to the small magnitude of the particle magnetic moment.  Magnetic potential interactions are quite small compared to thermal energy ($\sim$ 0.1 kT) and viscous forces between particles are quite large compared to magnetic forces (Mason number = 500)   \cite{critters}.  
Individual microrollers in the suspension are strongly coupled to the motion of their neighbors.  This is due to the rapid flows generated by rotating the microrollers close to a nearby floor \cite{Blake1974}.  The velocity of these collective flows is much higher than the individual translation velocity of an isolated microroller, a phenomenon that has been surprisingly overlooked until quite recently \cite{critters,Martinez2015}. It is these rapid flows that lead to the novel types of structures we study here.  In a homogeneous suspension of microrollers, this strong hydrodynamic coupling gives rise to a mean suspension translation velocity which increases linearly with \modif{the number density} $\rho_0$ \cite{critters}.  \modif{This is quite different from active systems of rolling particles, such as the Quincke rollers \cite{Bricard2013,Bricard2015}, where translational collective effects are much weaker, i.e.\ where the mean roller velocity weakly depends on $\rho_0$ (see SI Fig.1 in \cite{Bricard2015})}.   \modif{In the far-field, the flow generated by a microroller (or a Quincke roller) in the plane parallel to the wall (see Fig. 1c) has a faster decay ($\sim 1/r^3$) and a different structure from the dipolar flow field ($\sim 1/r^2$) observed in other systems of confined, driven suspensions of droplets or microswimmers \cite{Beatus2012,Brotto2013,Desreumaux2013}, see Fig.\ \ref{fig:rotlet_sketch}d.} This change is due to the difference in confinement (Hele-Shaw cell vs.\ a single boundary \cite{Blake1974, Hackborn1990}).

We create a uniform density  suspension by first mixing our sample, then loading it into the  chamber and letting the particles sediment to the chamber floor. The initial mixing ensures a uniform density profile, $\rho(x,y,t=0) \approx \rho_0$, across the chamber. Once the magnetic field is turned on, we observe transient density fluctuations: small clusters of particles which form and break up continuously (see Fig. \ref{fig:system_kymo}a and Movie1 in the SI). From the images, we extract the density fluctuations, $\delta\rho(x,y,t)  = \rho(x,y,t)-\rho_0$  \footnote{Working with high-density suspensions makes particle tracking challenging.  Therefore, we measure the intensity (e.g.\ `blackness') of the images, and use this as a proxy for density.}.
 The Fourier transform of these fluctuations $\delta\tilde{\rho}(\mathbf{k},\sigma) = 1/2\pi\int \delta\rho(\mathbf{r},t)e^{i(\mathbf{k}\cdot\mathbf{r}-\sigma t)}d\mathbf{r}$  is then used to  extract the pulsation $\omega^\prime(\mathbf{k})$, in a manner similar to that used in \cite{Desreumaux2013}, where $\mathbf{k} = (k_x = 2\pi/\lambda_x,k_y = 2\pi/\lambda_y)$.
Fig. \ref{fig:system_kymo}b shows the dispersion curve in the frame moving with the mean roller translational velocity $\langle V_x \rangle$: $\omega(\mathbf{k}) = \omega'(\mathbf{k}) - \langle V_x \rangle k_x$. 
Surprisingly, even though the particle-induced flow field is quite different from the dipolar one (Fig.\ \ref{fig:rotlet_sketch}c,d) observed in suspensions of strongly confined (q2D) particles \cite{Hackborn1990,Cui2004,  Desreumaux2013,Brotto2013}, the spectrum we measure is qualitatively similar: $\omega$ is symmetric about the axis $k_y = 0$ and antisymmetric about the axis $k_x = 0$. Density fluctuations propagate freely in all directions except for $k_x = 0$ and  their magnitude and direction of propagation change with $k_x$ and $k_y$. As shown on Fig.\ \ref{fig:rotlet_sketch}c,d, the rotlet and dipole flows share axial symmetry about the orientation axis (here $\bm{\hat{x}}$) and they are both attractive at the rear and repulsive at the front of the particle, which are the essential features needed to observe this propagative dynamics. 

To better understand this dispersive behavior, we introduce a minimal continuum model.  This model neglects out of plane motion in the $\bm{\hat{z}}$-direction.  We model the microroller suspension as an infinite sheet of rotlets (point-torque singularities). We consider a uniform plane of these rotlets  with planar density $\rho(x,y,t)$, which is fixed  at a height $z=h$. The value of $h$ used in the model is the gravitational height $h=a+k_BT/mg = 1\mu m$.  A point rotlet located at  $(x',y';\,h)$ induces a fluid velocity $\bm{v}(x,y;\,h)$  given by \cite{Blake1974}:
\begin{eqnarray}
v_{x}(x,y;\,h) &=& K_x\left(x-x',y-y';\,h\right) \nonumber\\
&=&  Sh\tfrac{(x-x')^2}{\left[(x-x')^2 + (y-y')^2 + 4h^2\right]^{5/2}} \label{eq:int_kernelx}
\end{eqnarray}
\begin{eqnarray}
v_{y}(x,y;\,h) &=& K_y\left(x-x',y-y';\,h\right) \nonumber\\
&= & Sh\tfrac{(x-x')(y-y')}{\left[(x-x')^2 + (y-y')^2 + 4h^2\right]^{5/2}}, \label{eq:int_kernely}
\end{eqnarray}
where $S = 6T_y/(8\pi\eta)$ in m$^3$/s,  $T_y = 8\pi\eta a^3 \Omega$  is the magnetic constant torque around the $y$-axis \footnote{In the experiments, the microrollers rotate synchronously with the magnetic field at a rate $\Omega$. In the simulations, we apply a constant torque, which is quantitatively similar to prescribing a constant rotation rate as shown in \cite{critters,Usabiaga2016b}.}
and $\eta = 10^{-3}$ Pa$\cdot$s is the dynamic viscosity of water.  The conservation law for the rotlet density in the sheet is
\begin{eqnarray}
\frac{\partial \rho(x,y,t)}{\partial t} &=& -\frac{\partial \left(\rho V_{y} \right)}{\partial y}  -\frac{\partial \left(\rho V_{x} \right)}{\partial x},
\label{eq:governing_2D}
\end{eqnarray}
where $V_{x}(x,y)$ and $V_{y}(x,y)$ are the local velocities due to nonlocal hydrodynamic interactions with rotlets at other positions, $
 V_{x}(x,y)  = K_{x}*\rho$ and
  $V_{y}(x,y) =    K_{y}*\rho$. 
We note that $V_x$ and $V_y$ are finite because the kernels $K_x$ and $K_y$ are not singular.\\ 
We linearize Eq.\ \eqref{eq:governing_2D}  about the uniform density state $\rho(x,y,t) = \rho_0 + \delta\rho(x,y,t)$, where $\delta\rho \ll \rho_0$, and get
\begin{eqnarray}
\frac{\partial \delta\rho}{\partial t} &=& -\rho_0\frac{\partial \left[ K_{y}*\delta\rho \right]}{\partial y} -\frac{\partial \left[\left\langle V_x\right\rangle \delta\rho + \rho_0 K_{x}*\delta\rho \right]}{\partial x},
\label{eq:governing_2D_linearized}
\end{eqnarray}
where $ \left\langle V_x\right\rangle = \rho_0 \int\limits_{-\infty}^{+\infty}\int\limits_{-\infty}^{+\infty} K_x(x,y;\,h) dx dy = \rho_0 \tfrac{\pi S}{3}$ is the theoretical mean roller velocity. 
We then seek plane wave solutions, $\delta\rho = \sum_{\mathbf{k}}\delta\tilde{\rho}_{\mathbf{k}}e^{i(\mathbf{k}\cdot\mathbf{x}-\sigma t)}$, of the linearized equation \eqref{eq:governing_2D_linearized} and extract the pulsation in the moving frame, $\omega(\mathbf{k}) = \omega'(\mathbf{k}) - \langle V_x \rangle k_x$, 
to obtain the following dispersion relation:
\begin{eqnarray}
\omega(k_x,k_y)  &=&  k_x\left\langle V_x\right\rangle\exp(-2hk)(1-2hk).
\label{eq:dispersion_2D}
\end{eqnarray}
The dispersion relation we find is purely real, which indicates that, in accordance with the experiments,  density waves freely propagate in the homogeneous suspension.  Additionally, the velocity and direction of propagation of these fluctuations is dependent on their wavelength, i.e. they are dispersive. 

The mean suspension velocity obtained from the continuum model $\left\langle V_x\right\rangle = \rho_0\frac{\pi S}{3}$ overestimates the one measured in the experiments by a factor of around 4-5. This is due to the fact that finite particle size effects and the suspension microstructure are not accounted for by the model. Therefore, to compare the speed of the density waves with the mean suspension velocity between theory and experiment, we normalize $\omega$  by $\left\langle V_x\right\rangle$.
The calculated dispersion curve \eqref{eq:dispersion_2D} is shown in Fig. \ref{fig:system_kymo}c and is in good qualitative agreement with the experimental dispersion curve shown in Fig.\ \ref{fig:system_kymo}b.

Fig.\ \ref{fig:system_kymo}d compares these two results  for $k_y=0$, $\phi_0 = 0.07 - 0.29$, where $\phi_0=\pi a^2\rho_0$ is the area fraction,  and driving frequencies $\Omega = 32.4-125.7$ rad/s. Due to the linear scaling of $\omega$ with $\left\langle V_x\right\rangle$, the experimental data sets can be collapsed with the rescaling $\omega h/\left\langle V_x\right\rangle$.
The data collapses well at longer wavelengths (above $\lambda_x>10h$). Below this value, the spread in the data is larger. We believe this is due to the fact that near field interactions such as contact forces become dominant and break the linear scaling at smaller wavelengths. 
The theoretical curve (Eq.\ \eqref{eq:dispersion_2D}) is in excellent agreement with the experimental results in the collapsed region and differs for smaller wavelengths. We argue that this departure is to be expected, since the continuum model neglects the near field steric and hydrodynamic interactions which become dominant at small length scales. We note that the dispersion relation is set only by $h$ and does not depend on the particle size at large $\lambda_x$ ($\lambda_x>10h$).  
These results demonstrate that far field hydrodynamic interactions drive traveling density waves in this system. 
Since the density fluctuations in a homogeneous suspension of microrollers are propagative, a suspension that is initially uniform will remain (on average) homogeneously distributed; at least within a linearized model. While the fully nonlinear model is not easily tractable theoretically, long-time numerical simulations and experiments have not indicated any apparent nonlinear growth of the fluctuations.

\begin{figure*}
 \includegraphics[width=1.9\columnwidth]{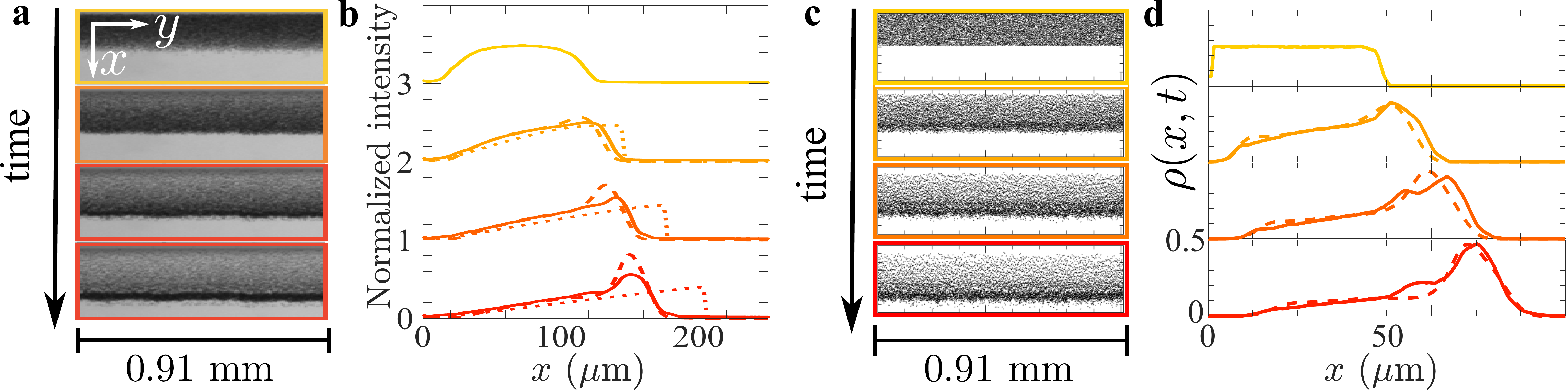}
\caption{ \textbf{a} Experiment: formation of the shock front with $\Omega = 125.7$ rad/s, images are 400 ms apart. \textbf{b} Solid line: normalized experimental intensity measurements for the images in \textbf{a}.  Each curve represents the mean intensity, with the mean taken  in the $\bm{\hat{y}}$-direction.  Dashed line:  nonlocal hydrodynamic model (\ref{eq:nonlocal_shock}). Dotted line: local model (\ref{eq:local_shock}). The curves are vertically displaced from each other for clarity. 
\textbf{c} Three dimensional particle simulation with $\Omega = 62.8$ rad/s, images are 640 ms apart. \textbf{d} Solid line: density,  $\rho(x,t)$, from the simulations averaged over four realizations. Dashed-line: nonlocal hydrodynamic model \eqref{eq:nonlocal_shock}.}
\label{fig:shock}
\end{figure*}

When sharp density gradients are introduced in the suspension, the response of the system is no longer purely propagative; the strong density gradient evolves into a traveling band, see Fig.\ \ref{fig:shock}a. In a second set of experiments, instead of a uniform distribution, we initially localize the particles  in a narrow strip on one side of the chamber.  After the rotating field is turned on, the particle distribution changes dramatically, organizing into a shock front, as shown in Fig. \ref{fig:shock}a and Movie2 in the SI.  
Shock-like structures have been observed in other driven suspension systems  \cite{Beatus2009, Champagne2011,Lefauve2014, Tsang2016}, where density shock waves are formed due to local   \cite{Champagne2011,Beatus2009} or nonlocal   \cite{Lefauve2014,Tsang2016} hydrodynamic interactions. However, in all of these cases, the shock evolves into a  Burgers-like shape; the shock front continually steepens and a  sharp discontinuity in density is observed.  Here, we observe something quite different: the shock in this microroller system evolves to have a finite width.  The curves in Fig.\ \ref{fig:shock}b represent the intensity measured in the propagating $\bm{\hat{x}}$-direction, where we have averaged over the transverse direction ($\bm{\hat{y}}$).  Our  measurements show that the shock front in this  system exhibits a well defined bump-like shape with a finite width. To understand the origin of this finite-width shock, we turn again to a continuum model, representing the microroller suspension as a continuum sheet of rotlets.

For short times, we can assume the rotlet density to be uniform along the $\bm{\hat{y}}$-direction (see Fig.\ \ref{fig:shock}a). 
Thus, after integrating  Eq.\ \eqref{eq:governing_2D} over the $\bm{\hat{y}}$-direction we obtain the one-dimensional \emph{nonlocal} conservation equation 
\begin{eqnarray}
\frac{\partial \rho}{\partial t} &=& -\frac{\partial \left[\rho \left(K*\rho\right) \right]}{\partial x},
\label{eq:nonlocal_shock}
\end{eqnarray}
where \modif{$$K = \int\limits_{-\infty}^{+\infty} K_x(x-x',y;\,h)dy = \frac{4Sh}{3}\frac{(x-x')^2}{\left[(x-x')^2 + 4h^2\right]^2}$$}. 
We note that, in the long wavelength limit, our \emph{nonlocal} model (Eq.\ \eqref{eq:nonlocal_shock}) \modif{can be approximated by the} \emph{local} inviscid Burgers' equation:
\begin{eqnarray}
\frac{\partial \rho(x,t)}{\partial t} &=& -\frac{\pi S}{3}\frac{\partial \rho(x,t)^2}{\partial x}.
\label{eq:local_shock}
\end{eqnarray}
We solved Eqs.\ \eqref{eq:nonlocal_shock} and \eqref{eq:local_shock}  numerically  with a standard finite volume solver. The initial conditions are taken from the normalized experimental intensity profiles. 
 When the shock forms, the particle height distribution, $P(h)$, is strongly modified from the equilibrium distribution. As a result, the average particle height, $h$, is greater than the gravitational height \cite{Usabiaga2016b}. Since $P(h)$ is difficult to measure experimentally, we  use an estimate for $h$ in our model.
Fig.\ \ref{fig:shock}b compares the numerical solutions of the local \eqref{eq:local_shock} and nonlocal \eqref{eq:nonlocal_shock}  equations with the experimental profiles. 
As seen in that figure, the nonlocal model accurately captures both the shape and the dynamics of the shock, although the magnitude of the bump is smaller in the experiments. This is due to our measurements underestimating density when microrollers pile on top of each other. 
In the model, $h$ is chosen so that the final front width matches the experimental one, $h = 2$ $\mu$m. This is consistent with the value we measure in particle-based 3D numerical simulations, as shown below.

Qualitatively different dynamics occur if the initial profile we measure is instead evolved according to a local  Burgers'  equation.  As shown in Fig.\ \ref{fig:shock}b, the contrast between the nonlocal  \eqref{eq:nonlocal_shock} and local \eqref{eq:local_shock} model is stark: the local model  does not capture the shape nor the evolution of the front  \cite{Beatus2009,Champagne2011}. 

To check the predictions of our nonlocal model quantitatively, we perform a direct comparison with particle-based Brownian dynamics 3D numerical simulations of the experimental system (so that the density $\rho$ and height $P(h)$ distributions are known exactly). Our simulation tool, described in \cite{Usabiaga2016b}, includes hydrodynamic interactions between the particles and between the particles and the floor, Brownian motion and steric interactions.  Each simulation contains $N= 32,768$ particles which are initialized by sampling the equilibrium Gibbs-Boltzmann distribution using a Monte Carlo method.  Each particle is subject to an external constant torque $T_y = 8\pi\eta a^3 \Omega$, where $\Omega = 62.8$ rad/s. 
Fig.\ \ref{fig:shock}c shows the time evolution of a portion of the suspension for $t = 0 - 2.56$ s. As in the experimental case, the initially homogeneous strip evolves into a shock region with a well defined width. 
Fig.\ \ref{fig:shock}d compares $\rho(x,t)$ from the particle simulations averaged over four realizations with the continuum model with no adjustable parameters. The height in the continuum model, $h = 2.62$ $\mu$m, is taken from the averaged particle height measured in the 3D simulations between $t=0$ s and $t=2.56$ s. The results shows that the continuum model is in quantitative agreement with the 3D simulations, thus confirming that the flow field in the $x-y$ plane at the average particle height plays a major role in the formation of shocks in these microroller suspensions.
Overall, these results demonstrate that the width of the shock front is intrinsically selected by the nonlocal nature of the hydrodynamic interactions, and  is set solely by the average height from the wall $h$ (see Appendix \ref{AppB} for additional results on front width selection).\\

Theoretical studies using the types of rotlet models proposed here can also qualitatively explain the hydrodynamic fingering instability observed after the formation of the shock, as we will present in future publications. In future work we will also explore with more detailed computational models the dependence of the mean velocity for a uniform suspension, which is not well-predicted by the simple theory presented here.

We expect similarly rich behavior dominated by hydrodynamic interactions with the floor and among particles in other systems where \modif{external boundaries} play a large role in determining the flow field. Our model, simulations and experiments can readily be extended to other systems, for example, the sedimentation of particles  adjacent to a wall. 
Although much work has been done to understand the dynamics and local structure of freely sedimenting particles  \cite{Guazzelli2011}, much less is understood about how nearby boundaries modify this system.  \\

This work was supported primarily by the Gordon and Betty Moore Foundation through Grant GBMF3849 and the Materials Research Science and Engineering Center (MRSEC) program of the National Science Foundation under Award Number DMR-1420073. P. Chaikin was partially supported by the Center for Bio-Inspired Energy Science, A DOE BES EFRC under Award DE-SC0000989. A. Donev and B. Delmotte were supported in part by the National Science Foundation under award DMS-1418706. We gratefully acknowledge the support of NVIDIA Corporation with the donation of GPU hardware for performing some of the simulations reported here.

\appendix

\section{Front width selection}
\label{AppB}
\begin{figure}
\includegraphics[width=0.9\columnwidth]{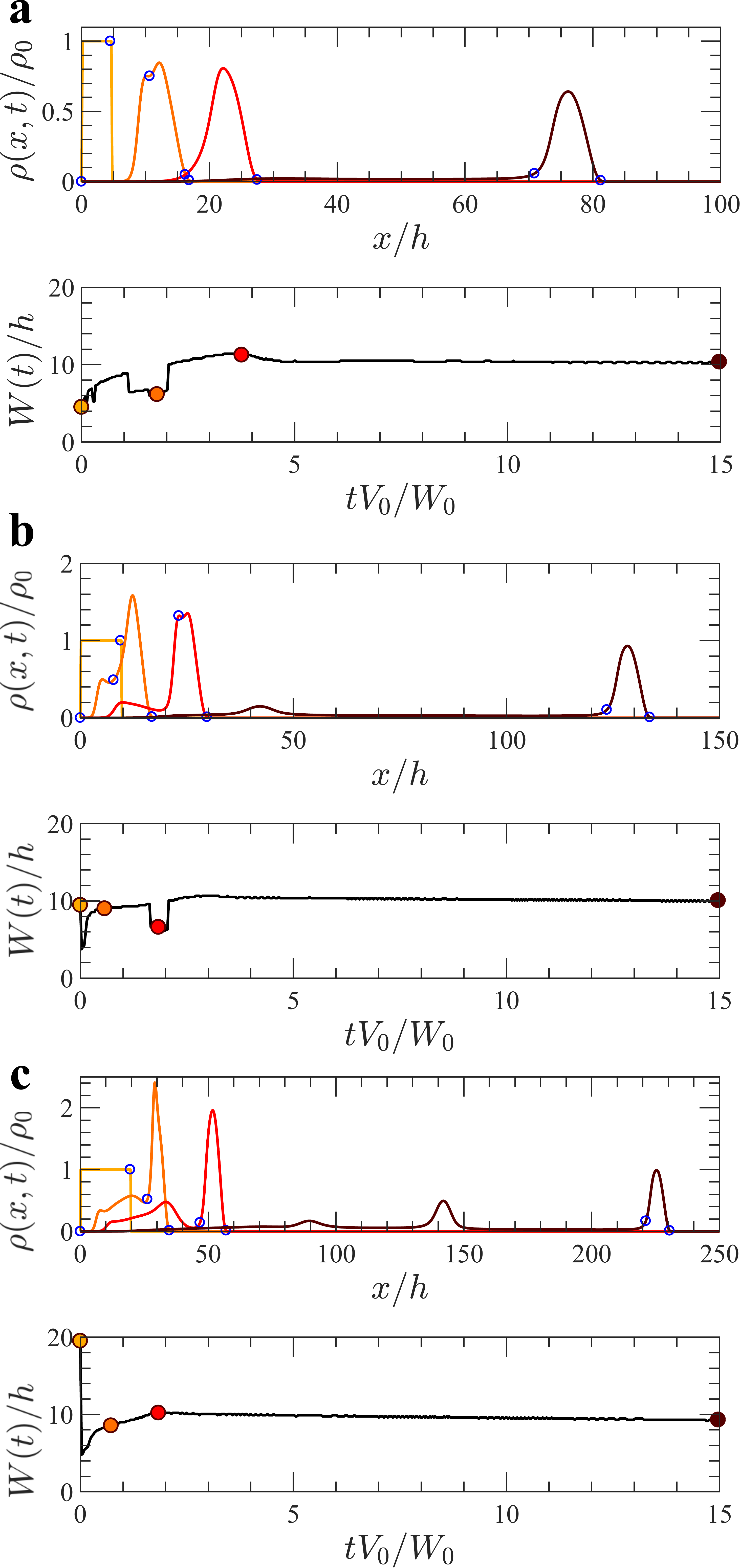} 
\caption{ Front width selection due to nonlocal hydrodynamic interactions. \textbf{a}
}  $W_0 = 5h$. \textbf{b}
  $W_0 = 10h$. \textbf{c}
  $W_0 = 20h$. Top: normalized density distribution. The color code from light orange to black represent increasing times. The blue circles delimit the front according to the criteria defined in the  text. Bottom: normalized front width vs.\ normalized time. The colored disks represent the times at which  $\rho(x,t)$ is shown in the top panel.
\label{fig:size_selection}

\end{figure}

To understand how the width of the shock front is selected, we numerically investigated the evolution of the nonlocal continuum model (Eq. (6) in the main text) at long times. 
We simulate  three different strips with initial widths: $W_0 = 5h, 10h, 20h$.  A fixed quantity of rotlets $M$ is initially uniformly distributed along the front:  $\rho_0 =  W_0/M $.
We monitor the front width $W(t)$ over time. $W(t)$ is defined as the  distance between the fore part where $\rho(x,t)<\epsilon_f = 0.025\rho_0$ and the aft  where  $\partial\rho(x,t)/\partial x  = \epsilon_a$, where  $\epsilon_a  = 10^{-2}$ $\mu$m$^{-2}$.   
Figure \ref{fig:size_selection} shows the time evolution of the density $\rho(x,t)$ and the front width $W(t)$ for $W_0 = 5h, 10h, 20h$. 
In this figure, time is normalized by the mean initial velocity of the front $V_0 = 1/M\int_0^{W_0}K(x)dx$, where $K(x)$ is defined in Eq (8) in the main text.

We observe that, independently of the initial width $W_0$, $W(t)$ quickly converges to a fixed value $W^{\star} \approx 10h$. The initially narrow strip $W_0 <W^{\star}$ spreads, while the wide one $W_0 > W^{\star}$ shed  particles and quickly divide to  reach $W^{\star}$; the system always evolves towards a shock-front with a width $\sim 10h$, regardless of the initial particle distribution.

\bibliographystyle{unsrt}

\bibliography{Hydrodynamic_shocks_in_microroller_suspensions_bib}

\end{document}